\documentclass[iop,apjfonts]{emulateapj}

\usepackage{hyperref}
\hypersetup{
colorlinks=true,
linkcolor=red,
citecolor=blue,
filecolor=cyan,
urlcolor=magenta,
}

\usepackage{graphicx}
\usepackage{amssymb}
\usepackage{amsmath}

\usepackage{multirow}
\usepackage{subfigure}
\usepackage{etoolbox}

\begin{document}

\title[Solar neutrino physics: Sensitivity to light dark matter particles]{
Solar neutrino physics: Sensitivity to light dark matter particles}
\author{Il\'\i dio Lopes~\altaffilmark{1,2} and Joe Silk~\altaffilmark{3,4,5}}
 
\altaffiltext{1}{Centro Multidisciplinar de Astrof\'{\i}sica, Instituto Superior T\'ecnico, 
Universidade Tecnica de Lisboa , Av. Rovisco Pais, 1049-001 Lisboa, Portugal;ilidio.lopes@ist.utl.pt,ilopes@uevora.pt } 
\altaffiltext{2}{Departamento de F\'\i sica,Escola de Ciencia e Tecnologia, 
Universidade de \'Evora, Col\'egio Luis Ant\'onio Verney, 7002-554 \'Evora - Portugal} 
\altaffiltext{3}{Institut d'Astrophysique de Paris, UMR 7095 CNRS, Universit\'e Pierre et Marie Curie, 98 bis Boulevard Arago, Paris 75014, France; silk@astro.ox.ac.uk}
\altaffiltext{4}{Department of Physics and Astronomy, 3701 San Martin Drive, The Johns Hopkins University, Baltimore MD 21218, USA}
\altaffiltext{5}{Beecroft Institute of Particle Astrophysics and Cosmology, 1 Keble Road, University of Oxford, Oxford OX1 3RH UK} 
\date{Received 2011 November 9; Accepted 2012 April 13}

\begin{abstract}
Neutrinos are produced in several neutrino nuclear reactions of the
proton--proton chain and  carbon--nitrogen--oxygen cycle that 
take place at different radius of the Sun's core.  
Hence,  measurements of solar neutrino fluxes provide a precise 
determination of the local temperature. 
The accumulation of non-annihilating light dark matter particles
(with masses between $5 \; {\rm GeV} $  and $16 \; {\rm GeV} $)  
in the Sun  produces  a change in the local solar structure, namely,
a decrease in the central temperature of a few percent.
This variation depends on the properties of the dark matter
particles, such as the mass of the particle and its spin-independent scattering cross-section on
baryon-nuclei, specifically, the scattering with helium, oxygen, and nitrogen 
among other heavy elements.
This temperature effect can be measured in almost all solar neutrino fluxes. 
In particular, by comparing the neutrino fluxes generated by  stellar models
with current observations, namely $^8$B neutrino fluxes,
we find that non-annihilating dark matter particles with 
a mass smaller than $10 \; {\rm GeV}$ and 
a spin-independent scattering cross-section with heavy baryon-nuclei larger than
$3 \times 10^{-37}\;{\rm cm^{-2}}$ produce
a variation in the $^8$B neutrino fluxes that would be in conflict with current measurements. 
\end{abstract}


\maketitle

\section{Introduction\label{sec-intro}}

$$\qquad$$
Dark matter makes up to 23\% of all the known matter of the universe. 
This evidence comes from cosmological
observations and numerical simulations~\citep{2011ApJS..192...18K}, 
which suggest that most of the data can only be explained by the presence of
a gravitational field caused by a new type of non-relativistic and non-baryonic fundamental particle \citep{Munshi:2011jg}.  
If such particles exist, then 
they should be observed in direct detection experiments, and their 
effects should affect the internal structure of large self-gravitating bodies 
such as the Earth and our own Sun. Thus, it is no surprise that so
much of the theoretical work  and experimental efforts 
in the fields of astrophysics, cosmology and particle physics have 
been dedicated to the discovery of such fundamental particles, 
hitherto with little success. 

Among the large number of dark matter particle candidates that have been proposed, 
Weakly Interacting Massive Particles (WIMPs) were until recently  among the most popular. 
WIMPs arise in several extensions of the Standard Model of fundamental particles,  
such as the supersymmetric models of fundamental particles \citep[e.g.,][]{1996PhR...267..195J}. 
In such classes of models, the lightest supersymmetric particle (LSP) , the neutralino, or possibly the NLSP, is the most 
suitable candidate for dark matter. The neutralino is the natural WIMP candidate, 
a stable particle with a self-annihilation cross-section 
of the order of the  weak-scale interaction. Unfortunately, 
this particle does not resolve the problem of  baryon-anti-baryon asymmetry. 
Recently, several authors \citep[e.g.,][]{2006PhRvD..73k5003G,2009PhRvD..80c7702F,2005PhLB..605..228H}
have suggested a new fundamental particle that carries a conserved charge 
analogous to the baryon number and would resolve the afore-mentioned problem \citep{2009PhRvD..79k5016K}.
This new type of matter, known as asymmetric dark matter and consisting mostly of Dirac particles, 
is particularly interesting since these candidates, 
like WIMPs, have interactions at  the  weak-scale and therefore 
sizeable scattering cross-sections with baryons, 
even if they do not annihilate  \citep{2009PhRvD..79k5016K}.
Asymmetric dark matter models \citep[e.g.,][]{2006PhRvL..96d1302F}  suggest that such particles 
should have a mass of the order of a few GeV \citep[e.g.,][]{2011arXiv1102.5644K,2010PhRvD..82e6001C}.
 
In recent years,  
several underground experiments  have been built  
to search for direct interactions of a dark matter particle with a baryon-nucleus
\citep[e.g.,][]{2005PhR...405..279B}. 
Amongst current results, positive indication for dark matter 
was produced by the DAMA/LIBRA\citep{2011PhRvD..84e5014B,2008EPJC...56..333B} 
and CoGeNT\citep{2011PhRvL.107n1301A} experimental teams.
Both of them have reported evidence of an annual modulation in the differential event rate,
which they explain as a consequence of the motion of the Earth around the Sun \citep{1986PhRvD..33.3495D}.
The simplest interpretation of the data is to assign the observed  annual modulation to be caused by  
the collision of a dark matter particle with a nucleus inside the detector.  
The dark matter particle is estimated to have a 
mass of the order of  a few GeV (likely between  $5$ and $12$ GeV),   and a
spin-independent scattering cross-section off baryons on the order of $10^{-40}\; {\rm cm^2}$.
Both experiments show identical positive detections, yet the DAMA experiment is 
favorable  to a larger scattering  cross-section than the CoGeNT experiment. 
Unfortunately,  other direct dark matter search experiments,  
XENON10/100 \citep{2011PhRvL.107e1301A,2011PhRvL.107m1302A} 
and CDMS \citep{2011PhRvL.106m1302A}  found null detections,
therefore these experimental results are in disagreement with the previous ones. 
Most recently, CRESST has also reported unexplained events in a similar mass range
\citep{2012PhRvD..85b1301B}.

There is an equally serious objection from  both cosmic microwave background and dwarf galaxy studies. 
WMAP7 uses acoustic peak damping to  constrain any annihilation contribution to early ionization \citep{2012JCAP...03..015C},
FERMI stacking of dwarf galaxies shows similar limits on the annihilation cross-section as a function of the mass
\citep{2011PhRvL.107x1303G}.
In the case of  thermal relic cross-sections, dark matter particles with masses below 10-30 GeV are excluded.
Nevertheless, several authors \citep[e.g.,][]{2010JCAP...08..018C,2011JCAP...11..010F,2011PhRvD..84h3001H,
2011PhRvD..84b7301D} 
have proposed various theoretical solutions to overcome the apparent inconsistencies in the data interpretation. 
Such theoretical proposals reconcile all the experimental results:  
DAMA/LIBRA and CoGeNT experiments are  no longer excluded by the CDMS and XENON  experiments; 
and the DAMA/LIBRA and CoGeNT experimental positive detections, previously,
leading to distinct and unrelated scattering cross-sections, now have the same values.
Likewise, the annihilation constraint is weakened if some fraction of the WIMPs do not annihilate, 
or  if the relic cross-section decouples from the current epoch cross-section by allowing for
s-wave suppression of the relic cross-section \citep{2011JCAP...07..003I}.
 
One of the more appealing suggestions is the possibility of
the dark matter particle being coupled  unequally to the nucleons (protons and neutrons) of the colliding  nuclei because of isospin violation. Usually, the  dark matter particle is considered to couple equally with protons and neutrons. 
Accordingly, the independent  scattering cross-section of heavy elements 
scales with $A^2$, where $A$ is the atomic number of the nucleus.
If the dark matter particles couple differently to protons and neutrons, these lead to a quite 
distinct interpretation of  direct dark matter searches.
This nuclear mechanism is known as isospin coupling violation 
\citep[e.g.,][]{2004PhRvD..69f3503K,2005PhRvL..95j1301G,2009NJPh...11j5026C}.
It was shown that a change of the strength of coupling of the dark matter particle
with protons and neutron allows the reconciliation of all of the current experiments  
\citep{2010JCAP...08..018C,2011JCAP...11..010F,2011PhRvD..84h3001H, 2011PhRvD..84b7301D}. 
In such cases, the proton scattering cross-section for the independent interaction increases 
by $10^2$--$10^3$ relative to the usual results,  leading to an effective independent scattering cross-section 
with values between $10^{-40} {\rm cm^2}$ and $10^{-36} {\rm cm^2}$ \citep{2011PhRvD..84b7301D,2011PhRvL.107n1301A,2011PhRvD..84h3001H}.  
We point out here that these types of dark matter particles can modify the structure of a star like the Sun, 
leading to a quite different flux of solar neutrinos.  
 
The measurement of the solar neutrino fluxes offers a distinct method 
for a diagnostic of the central regions of the Sun. In past decades, the main goal of solar neutrino detection 
was to study the particle properties of the neutrinos themselves. Now that neutrino oscillations are well-established, 
detectors are arriving at a stage where information on the solar core can be extracted from precise 
measurements of the solar neutrino spectrum. This enables  a search for dark matter in the core of the Sun.

In the next section, we review the main physical processes related with the capture of dark matter by a star
like the Sun. In Section 3, we present a detailed discussion about the production of neutrinos in the solar  core
and explain the impact of dark matter on the production of neutrino fluxes. In the final section we 
discuss the results obtained relative to the current experimental bounds.

\begin{figure}[!t]
\centering
\includegraphics[scale=0.45]{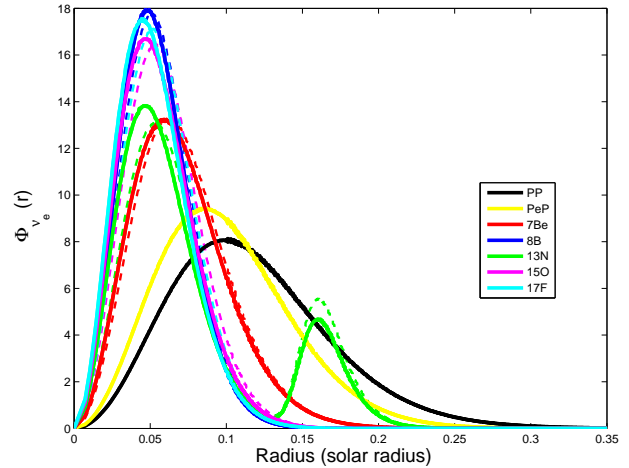}
\caption{The electron-neutrino flux produced in the various nuclear reactions of 
 the  PP chains and CNO cycle for two distinct solar models.
(i) continuous curves: the solar model is in agreement with the most current helioseismology diagnostic, 
 and other solar standard models published in the literature \cite{2005ApJ...621L..85B,2010ApJ...713.1108G}.
(ii) dotted curves: a solar model evolved in a halo of dark matter particles with mass of 
$7$ GeV and  independent scattering cross-section $\sigma_{\rm SI}=10^{-36} \;{\rm cm^2}$
(other parameters see text).  The neutrino flux produced in the various nuclear reactions  
is  computed for a revised  version of the solar model and the most updated microscopic physics data.}
\label{fig-neut1}
\end{figure}

\section{Sun's evolution in a non-annihilating  dark matter particle halo}  

\begin{figure}[!t]
 \centering
\includegraphics[scale=0.45]{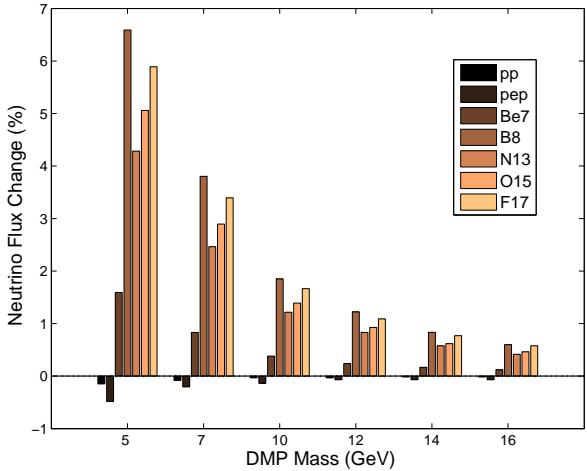}
\caption{Percentage  decrease  changes in the solar neutrino flux,  
$(\Phi_{ssm}-\Phi_{DM})/\Phi_{ssm}$. 
$\Phi_{DM}$ is one of the neutrino fluxes  for the PP chain ($pp$, $pep$, $^7$Be and $^8$B) or the CNO cycle
 ($^{13}$N, $^{15}$O and $^{17}$F ) in the case of non-annihilating dark matter particles accumulate in the center of the Sun. The   dark matter particles scatter with nucleons with $\sigma_{\rm SD}=10^{-40}\;{\rm cm^2}$
and $\sigma_{\rm SI}=10^{-36}\;{\rm cm^2}$. The mass of the dark matter particle is indicated in the figure.
The neutrino flux percentage change of each species is compared to
the current flux values as predicted by our solar standard model.}
 \label{fig-neut2}
 \end{figure}

The formation and evolution of the Sun inside a dark matter halo
is similar to a normal star of population II, although 
there are some important modifications \citep{2008dmap.conf..387S, 2009ApJ...705..135C,2009MNRAS.394...82S,2011PhRvD..83f3521L}.
The Sun by means of  its large gravitational field captures an
important amount of dark matter particles during its evolution
\citep{2010PhRvD..82j3503C,2010PhRvD..82h3509T}.
The accumulation  of dark matter inside the star leads to 
the formation of a dark matter core.  
The effect on the structure  of the star is primarily to provide 
an additional mechanism for the transfer of energy, 
which in some cases significantly reduces the temperature gradient, 
thereby causing the stellar core to become almost isothermal \citep{2002PhRvL..88o1303L,2010ApJ...722L..95L}.
In some dark matter scenarios, dark matter  annihilation 
provides the star with an extra source of energy which changes the evolution
path of the star \citep{2011ApJ...733L..51C,2011MNRAS.410..535C}. 
Nevertheless, such physical processes are not relevant for the dark matter models 
discussed here.  

The efficiency of the energy transport provided by dark matter particles
depends on the average distance traveled by  particles between 
consecutive collisions, i.e., the mean free path of the dark matter particle. 
If the mean free path is short compared with the dark matter scale height, 
then the energy transfer proceeds by conduction. Alternatively,  
if the mean free path is large, the successive collisions are widely separated,
so that the energy transfer proceeds within the Knudsen regime \citep{2002PhRvL..88o1303L}. 
In most of the dark matter scenarios  
discussed in this paper, the energy transport is dominated by conduction, 
although, the code has an implementation of  the transport of energy by dark matter 
in both regimes \citep{2002MNRAS.331..361L,2011PhRvD..83f3521L}. 

The amount of dark matter captured by the star, among other parameters,
depends on the mass of the dark matter particles $m_\chi$ and the 
collision cross-section of dark matter particles with baryons \citep{1987ApJ...321..571G}.
Two leading parameters define the scattering of the dark matter
particles with the baryon nuclei: the spin-dependent scattering cross-section $\sigma_{\rm SD}$ that
is only relevant for hydrogen nuclei; and the spin-independent scattering cross-section $\sigma_{\rm SI}$
that defines the interaction of the dark matter particles with the heavy nuclei.  The 
values of $\sigma_{\rm SI}$ used in the calculation include the values found by the
new interpretation of experimental results,
comprised by the DAMA/LIBRA and CoGeNT data.
If the value of  $\sigma_{\rm SI}$ is larger or equal to  $\sigma_{\rm SD}$,
the capture of dark matter particles is dominated by the collisions 
with heavy nuclei, rather than by  collisions with hydrogen \citep{2011PhRvD..83f3521L}.

The capture rate is computed numerically from the integral  expression of \citet{1987ApJ...321..571G} 
implemented as indicated in \citet{2004JCAP...07..008G}. 
If not stated otherwise,  we assume that the density of dark matter in the 
halo is $0.3 \; {\rm GeV cm^{−3}}$, the stellar velocity of the Sun is $220 \; {\rm km s^{-1}}$ 
and the Maxwellian velocity dispersion  is $270 \; {\rm km s^{-1}}$. 
The stellar code explicitly follows the capture rate of the dark matter particles by 
the different chemical elements present inside the Sun, some of them changing  
in  isotopic abundance during the Sun's evolution. 
The paper~\citep{2011PhRvD..83f3521L} presents  
a detailed discussion about the properties of  dark matter particles, as well 
as the impact that the uncertainties of the dark matter parameters  have 
on the evolution of the Sun and stars.

Our  evolution code is a modified version of the  one-dimensional stellar evolution CESAM
\citep{1997A&AS..124..597M}.  The  code has up-to-date  and very refined microscopic physics
(updated equation of state, opacities, nuclear reactions rates, 
and an accurate treatment of microscopic diffusion of heavy elements), 
including the solar mixture of \citet{Asplund:2005uv}.
The solar models are calibrated to the present solar radius $R_\odot= 6.9599 \times 10^{10} \;{\rm cm}$, 
luminosity  $L_\odot = 3.846 \times 10^{33} \; {\rm erg\; s}^{−1}$, mass $M_\odot = 1.989 \times 10^{33} \;{\rm g}$, 
and age $t_\odot = 4.54\pm 0.04\; {\rm Gyr}$  \citep[e.g.,][]{2011RPPh...74h6901T}. 
The models are required to have a fixed value of the photospheric ratio $(Z/X)_\odot$, where  
 $X$ and $Z$ are the mass fraction of hydrogen and the mass fraction of elements heavier than helium.
The value of $(Z/X)_\odot$ is determined according to the solar mixture proposed by \citet{Asplund:2005uv}.
Our reference model is a solar standard model \citep{1993ApJ...408..347T} that shows  
acoustic seismic diagnostics and solar neutrino fluxes identical  to other solar standard models 
\citep{2010ApJ...713.1108G,2009ApJ...705L.123S,2005ApJ...621L..85B,2010ApJ...715.1539T}.

We have computed the evolution of the Sun in
several dark matter particle scenarios,  where
$m_\chi$  takes values  between $5$ and $16$ GeV, 
and the spin-independent  scattering cross-section $\sigma_{\rm SD}$ 
takes values  between
$10^{-41}$ and $10^{-34}\;{\rm cm^2}$. The
spin-dependent scattering cross-section $\sigma_{\rm SD}$ is approximately $10^{-40}\;{\rm cm^2}$.  
This choice of parameters covers the parameter range of the experimental results of the direct 
dark matter searches discussed in the previous section.

The  Sun evolves from the beginning of the pre-main sequence
until its present age. Each solar model has more than 2000 layers,
and it takes more than 80 time steps to arrive at the present age.
In the scenarios where a large amount of dark
matter is captured, a single solar model can have more than 
140 time steps before arriving at the present Sun. 
 At each epoch, the structure equations  are solved
with an accuracy of $10^{-5}$. For each set of dark matter parameters, a solar-calibrated model is obtained by
automatically adjusting the helium abundance and the convection mixing length parameter until the total luminosity
and the solar radius are within $10^{-5}$ of the present solar values. Typically, a calibrated solar model
is obtained after a sequence of  10 intermediate solar models, although the models with a large concentration of dark matter 
need more than 20 intermediate models.  In the next section, we discuss the impact of such models on
the production of solar neutrino fluxes.

\begin{figure}[!t]
\centering
\includegraphics[scale=0.65]{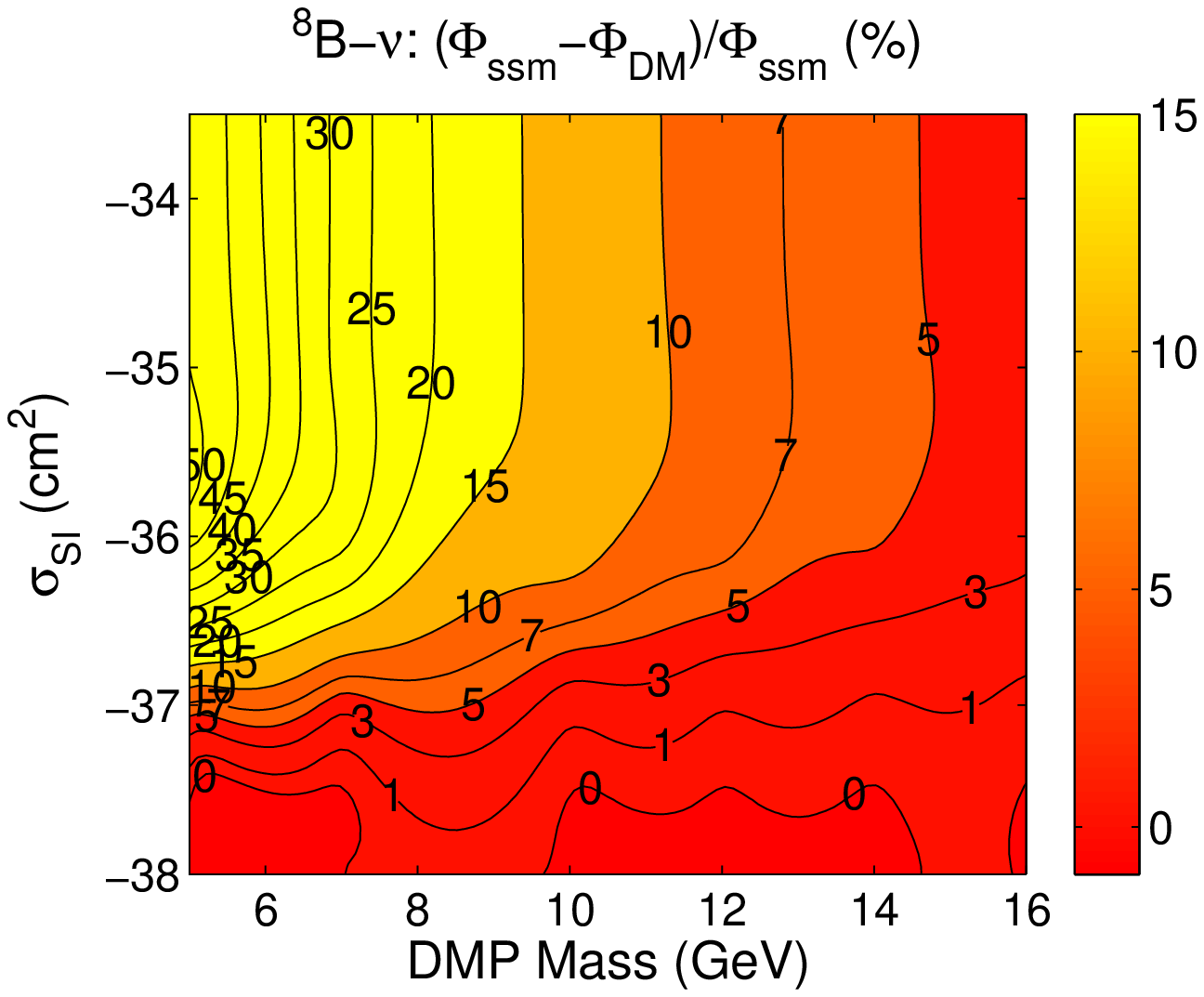} 
\includegraphics[scale=0.65]{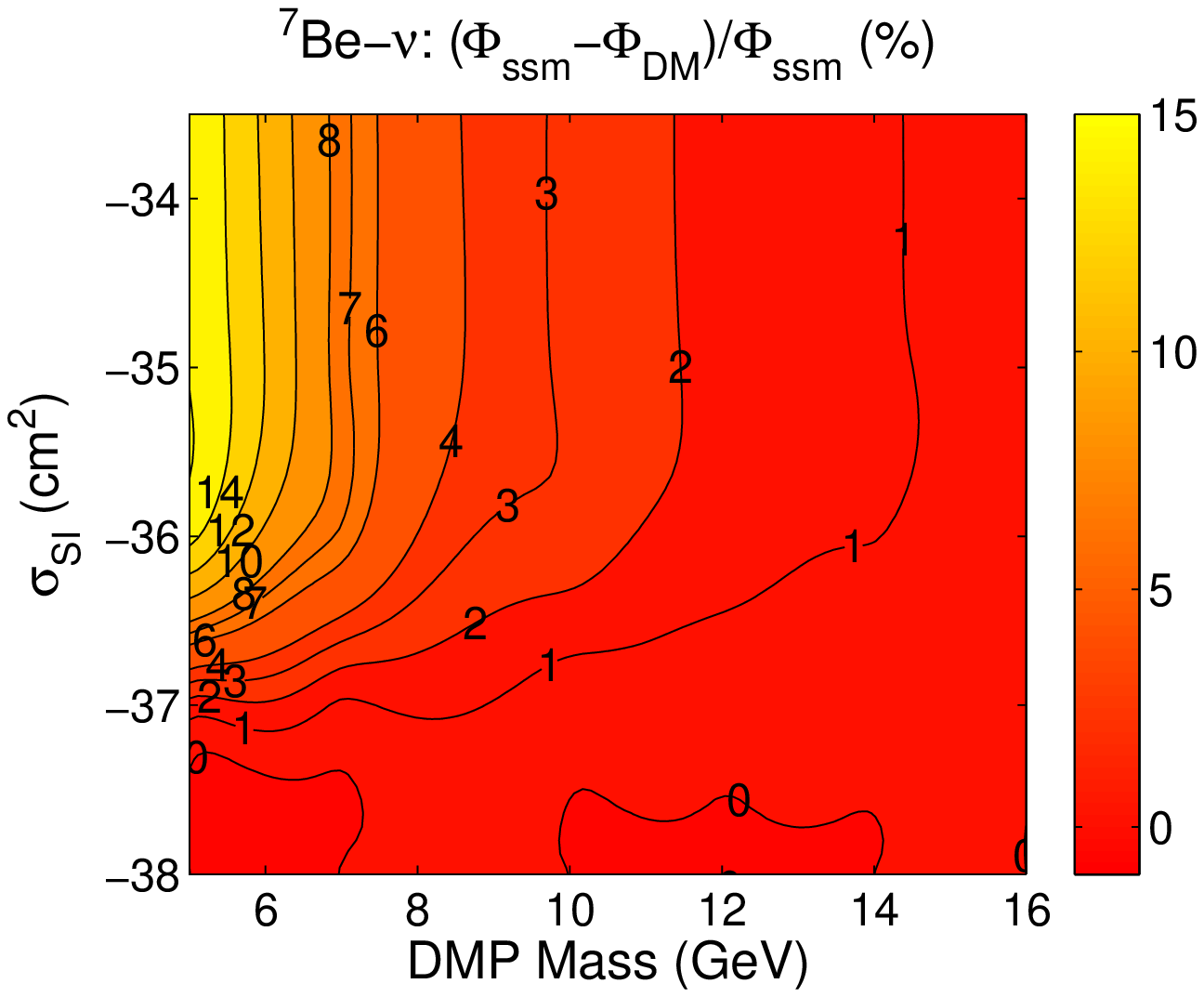}
\caption{Percentage decrease  changes in the solar neutrino flux of $\Phi_\nu(^8{\rm B})$ and 
 $\Phi_\nu(^7{\rm Be})$ computed as  $(\Phi_{ssm}-\Phi_{DM})/\Phi_{ssm}$.  
$\Phi_{DM}$ is one of the following fluxes $\Phi_\nu(^8{\rm B})$ (top figure) and  $\Phi_\nu(^7{\rm Be})$ (bottom figure).
Iso-curves for neutrino fluxes in the plane of the dark matter particle 
 independent scattering cross-section ($\sigma_{\rm SI}$) vs. 
 the dark matter particle mass ($m_\chi$). The other parameters of the
 dark matter particles and the stellar physics can be found in the main text. 
The neutrino fluxes of  the solar standard model are used as reference.
Both panels use the same contour colour scheme.}
\label{fig-neut3}
\end{figure}

\section{Solar neutrinos and dark matter}

Neutrinos are generated in the interior of the Sun 
as a by-product of the fusion processes that power the solar energy production, 
mainly by the proton-proton (PP) fusion chain (98\%), 
and with a small contribution from the catalytic carbon--nitrogen--oxygen (CNO) cycle (2\%).
Accordingly to the modern theory of neutrino oscillations \citep{2010arXiv1010.4131H},
solar neutrinos fluctuate between three possible flavors during  their travel throughout space.
Although solar neutrinos are generated in the Sun's core with electron flavor, 
depending on their energy, a significant part of them are converted into another flavor. 
The conversion from electron flavor is accomplished by 
means of two possible mechanisms:  
the neutrino vacuum oscillations and the
Mikheyev--Smirnov--Wolfenstein (MSW) neutrino oscillations or matter neutrino oscillations 
\citep{2008PhR...460....1G,2011PhRvD..84e3007F,1999PhLB..466..415A}.  

The change of structure of the solar core, caused by the presence of dark matter,
leads to  variations of the temperature and density profiles, which in turn produce 
a visible variation in the production of neutrino fluxes
\citep{2010PhRvD..82h3509T,2010PhRvL.105a1301F,2010PhRvD..82j3503C}.
This variation in the neutrino fluxes 
is mainly caused by a change in the rate of the production of neutrinos, due to the
decrease of the local temperature. This process is accompanied by a small increase of
the density profile, for which the impact on the MSW neutrino oscillation mechanism 
is negligible. For this reason, we will restrict our analysis to the total 
neutrino flux production and neglect the contribution to neutrino oscillations.

In the Sun's core, the neutrino emission regions occur in a sequence of shells, 
following closely  the location of nuclear reactions, orderly arranged in a 
sequence of decreasing temperatures, from the center towards the surface.
In the case of our solar standard model, the maximum neutrino emission of the 
 PP  nuclear reactions,  $pp$, $pep$,
$^7$Be and $^8$B, occurs at 10\%, 8\%, 6\% and 4\% of the solar radius. 
Figure~\ref{fig-neut1} shows the location of the different neutrino emission regions for several 
nuclear reactions.

The presence of  dark matter particles inside the star produces  
significant modifications of the solar neutrino fluxes (see Figure~\ref{fig-neut2}), 
caused by the reduction of the temperature and by changes in the neutrino emission regions 
(see Figure~\ref{fig-neut1}). The  $pp$ and $pep$ nuclear reactions have neutrino emission regions that  
extend from the center up to 30\% and 25\% of the solar radius.  
Both nuclear reactions are strongly dependent on the total luminosity of the star. 
As a consequence, different calibrated solar models (with or without dark matter) maintain the same 
neutrino emission shells. In Figure~\ref{fig-neut1}, the $pp$ and  $pep$ emission shells
are identical for both solar models. 
Instead, the $^8$B and $^7$Be neutrino emission shells with a  
width of 15\% and 22\% of the solar radius have slightly changed  the location
of their maxima. 
This effect has a strong impact on the neutrino fluxes from these
nuclear reactions  (see Figure~\ref{fig-neut2}), since the  $\Phi_\nu(^8{\rm B})$ neutrino flux dependence on the central temperature $T_c$ 
is proportional to $T_c^{24}$ and  $\Phi_\nu(^7{\rm Be})$ is proportional 
to $T_c^{10}$ \citep{1993ApJ...408..347T,1996PhRvD..53.4202B,2002PhRvC..65b5801B}.  
Likewise, the $\Phi_\nu(pep)$   is proportional to $T_c^{-2.4}$ \citep{1996PhRvD..53.4202B}.
It follows that $\Phi_\nu (^8{\rm B})$ is a more sensitive probe of the central temperature than $\Phi_\nu(^7{\rm Be})$.

The neutrino emission shell of the CNO cycle nuclear reactions,   
$^{15}$O, $^{17}$F and $^{13}$N has a  behavior identical
to that of the $^8$B neutrino emission shell. The only exception is the $^{13}$N nuclear reaction
that has a second emission shell located between 12\% and 25\% of the Sun's radius,
with a maximum emission at 16\% of the solar radius. The neutrino fluxes due to  
such nuclear reactions, similar to $^{8}$B, are very sensitive to the presence of dark matter 
in the Sun's core. 
 
Figure~\ref{fig-neut2} shows the change of neutrino fluxes relative to the solar standard
model for several dark matter scenarios of light dark matter particles.
The neutrinos produced in the central regions are more affected
than the ones produced in the more external layers. 
Furthermore, the  dark matter particles with smaller masses
produce a more extended  dark matter core leading to
a larger effect on the production of neutrino fluxes.


\section{Discussion and Conclusion}

\begin{figure}[!t]
\centering 
\includegraphics[scale=0.65]{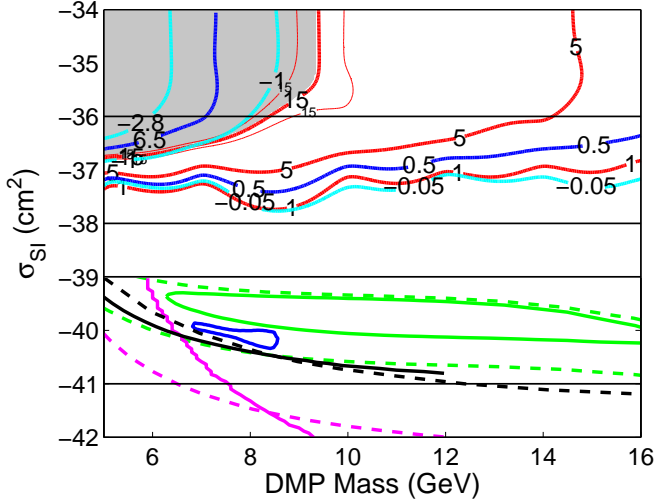}
\caption{Exclusion Plot $m_\chi $ -  $\sigma_{\rm SI}$ for current dark matter searches 
and future neutrino experiments :
{\bf (i)} $-$ Current experimental allowed regions and  exclusion contours:
CoGeNT Annual Modulation ROI  \citep{2011PhRvL.107n1301A}(closed  blue curve), 
and  DAMA/LIBRA assuming no ion channeling \citep{2009JCAP...04..010S,2008EPJC...56..333B}
with a $3\sigma$ CL  (solid green curve) and $5\sigma$ CL (dashed green curve);
the XENON 10 ($90\%CL$) \citep{2011PhRvL.107e1301A}  (dashed magenta curve)
and XENON100 \citep{2011PhRvL.107m1302A}  (solid magenta curve) exclusion contours; 
the  CDMS II limit \citep{2011PhRvL.106m1302A}  (solid black curve);
the  SIMPLE  limit \citep{2012PhRvL.108t1302F} (dashed black curve).
{\bf (ii)} $-$ iso-contour of solar neutrino fluxes expected percentage variations 
($(\Phi_{ssm}-\Phi_{DM})/\Phi_{ssm}$ ) for dark matter solar models.
$\Phi_{DM}$ is one of the following fluxes: 
$\Phi_\nu(^8{\rm B})$  -  solid red curve $15\%,5\%,1\% $,   
 $\Phi_\nu(^7{\rm Be})$ - solid blue curve $6.5\%,0.5\%$   and 
 $\Phi_\nu(pep)$ - solid cyan curve $-2.8\%,-1.0\%,-0.05\%$.
{\bf (iii)} $-$  The gray region corresponds to values of  $\Phi_\nu(^8{\rm B})$ 
neutrino flux larger than  $15\%$ (red curve), the  two  $15\%$-thin-red lines 
correspond to assuming an experimental error of $2\%$ on the measurement of $\Phi_\nu(^8{\rm B})$
(LENA experiment targets a 1\% error bar). 
{\bf (iv)} $-$ The region between the continuous black straight-lines 
corresponds to the location of  the theoretical dark matter candidates:  
experimental data is contradictory when dark matter interaction is interpreted assuming  the 
conservation of iso-spin ($\sigma_{\rm SI} $ with values $10^{-41}-10^{-39} {\rm cm^2}$ ),  and
all the dark  experimental data is reconciled when dark matter interaction is interpreted assuming
iso-spin violation,  inelastic scattering or some other similar theoretical mechanism 
 ($\sigma_{\rm SI} $ with values $10^{-38}- 10^{-36} {\rm cm^2}$).
}
\label{fig-neut4}
\end{figure}

The  $\Phi_\nu(^8{\rm B})$ and $\Phi_\nu(^7{\rm Be})$  neutrino
fluxes are presented as a variation relative to the values of the 
neutrino fluxes in the Solar Standard Model. 
This procedure facilitates the comparison
of these results with the data from the Sudbury Neutrino Observatory (SNO) and Borexino detectors.

The current neutrino experiments suggests that the expected value
for  the $^8$B neutrino flux in  the case of no neutrino oscillations (or electron neutrinos; including the theoretical uncertainty
on the $^8$B flux in the solar standard model) is  
$\Phi_\nu(^8{\rm B})= 5.05^{+0.19}_{-0.20} \times 10^6 {\rm cm^{−2} s^{−1}}$
for the SNO experiment \citep{2010PhRvC..81e5504A}, and
$\Phi_\nu (^8{\rm B})= 5.88\pm {0.65} \times 10^6 {\rm cm^{−2} s^{−1}}$ for the Borexino experiment
\citep{2010PhRvD..82c3006B}. 
Furthermore,  the Borexino experiment  measures $^7$Be solar neutrino rates with an accuracy better than $5\%$. This corresponds to a $^7$Be  neutrino flux  $\Phi_\nu(^7{\rm Be})= 4.87\pm 0.24 \times 10^9 {\rm cm^{−2} s^{−1}}$,
under the assumption of the MSW-LMA scenario of solar neutrino oscillations
\citep{2011PhRvL.107n1302B,2008PhLB..658..101B}. 
These values are in agreement with the neutrino predictions 
of the current solar standard model  \citep{2010ApJ...715.1539T,2011ApJ...743...24S}.
Nevertheless, the Sun's core seems to be slightly hotter than expected, since the $\Phi(^8{\rm B})$ 
value measured by Borexino and  SNO is higher than the value predicted by the 
current  solar standard model. 
This discrepancy is larger in the case of Borexino than for the SNO measurement. 
Furthermore, this result is validated by the Borexino measurement of $\Phi_\nu(^7{\rm Be}) $ which is 
sensitive to a  region  slightly off the Sun's  center (see Figure~\ref{fig-neut1}).

Recently, the Borexino experiment made the first measurement of the  $pep$ neutrinos, 
$\Phi_\nu(pep)= 1.6\pm 0.3 \times 10^8 {\rm cm^{−2} s^{−1}}$ \citep{2012PhRvL.108e1302B}.
This preliminary measurement of  $\Phi_\nu(pep)$ also indicates that the Sun is hotter than the solar standard model. 
This diagnostic is very interesting because $\Phi_\nu(pep)$ is strongly dependent on the luminosity of the star. 
Therefore, it is an indirect measurement of the total energy (luminosity) produced in the nuclear region, 
without the complications related with the transport of energy by radiation and convection 
of the more external layers of the star. Once the accuracy is improved in such measurements, 
this will be a powerful test to the nuclear region of solar models.

Dark matter particles with masses smaller than
5 GeV are not considered because evaporation becomes important and a large number of
dark matter particles then escape the gravitational field of the star, significantly  reducing the impact 
on the Sun's core \citep{1990ApJ...356..302G}. Dark matter particles with masses above $12$ GeV produce a
very small dark matter core and their effect in the Sun's structure diminishes rapidly, 
leading to  variations of at most a  few percent on the predicted neutrino fluxes \citep{1985ApJ...294..663S}.
 Moreover, for the light dark matter particles with masses between 5 and 12 GeV, 
the size of the affected core region is comparable to that of the neutrino generation region, 
thus reducing the neutrino count expected to be measured in the  terrestrial neutrino detectors.  

Figure~\ref{fig-neut3} shows a systematic study of the dark matter parameter space by
varying the mass $m_\chi$ and the spin-independent scattering cross-section 
$\sigma_{\rm SI}$ of the dark matter particles. A set of more than a hundred solar 
models was computed for different dark matter parameters. 
These results are qualitatively consistent with previous results
in the case of the spin-dependent scattering cross-section \citep{2010Sci...330..462L,2010ApJ...722L..95L}. 
In the $m_\chi$ vs. $\sigma_{\rm SI}$ plane, the iso-contours 
corresponding to $\Phi_\nu(^8{\rm B})$ and $\Phi_\nu(^7{\rm Be})$ show variations up to 50\% and 30\% 
relatively to the solar standard model.  The $\Phi_\nu(^8{\rm B})$ variation is larger than $\Phi_\nu(^7{\rm Be})$,  
because $^8$B neutrinos are produced in a more central region than  $^7$Be neutrinos, and
therefore have a larger sensitivity (see Figure~\ref{fig-neut2}).

The current modeling of the microscopic physics (nuclear reactions rates, microscopic diffusion, radiative transfer
and elemental abundances) in the Sun presents some significant uncertainties in the neutrino fluxes and
acoustic mode predictions. Nevertheless, several authors \citep{2005ApJ...621L..85B,2010ApJ...715.1539T,2010ApJ...713.1108G} 
have shown that  the largest contribution to the  theoretical uncertainties comes from new measurements 
of the surface elemental  abundances \citep{2009ARA&A..47..481A,2009ApJ...705L.123S}.
Indeed, the scientific community agrees that the new abundance determinations
are the most reliable. 
If we consider the combined experimental and theoretical uncertainties of $\Phi_\nu(^8{\rm B})$ to be of the order of 15\%, i.e.  
$\Delta \Phi_\nu(^8{\rm B})/\Phi_\nu(^8{\rm B})\sim 15\% $,  then the dark matter particles  for which the variation of 
$\Phi_\nu(^8{\rm B})$  is larger than 15\% can be excluded.
This corresponds approximately to rejecting dark matter particle candidates  with low mass and large scattering cross-sections, i.e., $ m_\chi\le 11\; {\rm GeV }$  and $ \sigma_{\rm SD}\ge 3\times 10^{-37}\;{\rm cm^2} $.
Moreover, the $^8$B flux variation  is equivalent to a variation of the central temperature,
$\Delta T_c/T_c $ to be of $0.6\%$.  Assuming that the other neutrino flux sources are
affected by the same uncertainty, the dark matter candidates excluded
have $ ^7Be$ and $pep$ neutrino fluxes, such as $\Delta \Phi_\nu(^7{\rm Be})/\Phi_\nu(^7{\rm Be})\ge 6.25\% $ or
$\Delta \Phi_\nu(pep)/\Phi_\nu(pep)\le - 1.5\% $.  

Figure~\ref{fig-neut4} shows the allowed regions and exclusion contours of  current dark matter searches
and the solar neutrino flux variations of several dark matter-modified solar models.  
The current positive detections of dark matter experiments 
such as  DAMA/LIBRA  and CoGeNT occur for values of  $\sigma_{\rm SI}$ that do not affect the evolution
of the Sun, $10^{-41}\;{\rm cm^2} \le \sigma_{\rm SI}\le 3\times 10^{-39}\;{\rm cm^2}$. 
This situation has changed with  a new interpretation of the experimental data results, 
which evokes  the possibility that if  dark matter interactions with ordinary matter do not conserve  iso-spin, as
several authors have showed,  then all the direct detection experimental data can be reconciled 
\citep[e.g.,][]{2011JCAP...11..010F}.  
Other processes such as   inelastic scatterings and channeling, 
velocity-dependent form factors and momentum-dependent form factors 
are among the various processes that may significantly increase the value of the spin-independent 
scattering cross-section of the dark matter particles
\citep[e.g.,][]{2011JCAP...11..010F}.  
Such new models suggest that the dark matter particle candidates have 
$\sigma_{\rm SI}$ values within reach of  the solar neutrino experiments, 
$10^{-38}\;{\rm cm^2} \le \sigma_{\rm SI}\le 3\times 10^{-36}\;{\rm cm^2}$.  

In the near future with the improvement of the sensitivity of the detectors,
it will be possible to reduce the uncertainties in the  $\Phi_\nu(^8{\rm B})$ , 
$\Phi_\nu(^7{\rm Be})$ and $\Phi_\nu(pep)$,
as well as to measure the neutrino fluxes produced in the CNO cycle. 

The CNO neutrino fluxes can provide an additional test of the accretion
of dark matter in the Sun's core, once the  $\Phi_\nu(^{13}{\rm N})$, $\Phi_\nu(^{15}{\rm O})$ and $\Phi_\nu(^{17}{\rm F})$ 
electronic neutrino fluxes have a dependence on $T_c$ of $24.4$, $27.1$ and $27.8$, identical to or larger
than the neutrino fluxes of  $^8$B  \citep{1996PhRvD..53.4202B}.
The high-statistics data expected to be collected by the upgraded versions of present
neutrino detectors, such as the SNO and  Borexino, 
or even more so by the new neutrino detector Low Energy Neutrino Astrophysics (LENA),
would allow a precise determination of such solar neutrino fluxes.
It is expected that  LENA will allow  measurement of  $^8$B neutrinos with a precision of 1\%
\citep{2011PhRvD..83c2010W}.  As illustrated in Figure~\ref{fig-neut4}, if an error bar of 2\%  in the measurement
of $\Phi_\nu(^8{\rm B})$ is obtained at the experimental level, this will allow us to infer 
an even more  significant  constraint  on the parameters of possible dark matter particles. 

Such detailed determinations of neutrino fluxes by future neutrino experiments will provide the means to 
precisely probe the central Sun's core temperature,  and by doing so, the contents of dark matter.
In particular, this will allow a better screening of dark matter particle candidates 
such as those postulated to fit the data of the DAMA/LIBRA  and CoGeNT experiments.
 
\begin{acknowledgments}
This work was supported in part by grants from "Funda\c c\~ao para a Ci\^encia e Tecnologia" (SFRH/BD/44321/2008),  "Funda\c c\~ao Calouste Gulbenkian", and  the NSF (grant OIA-1124453). 
We thank the anonymous referee for the useful comments and suggestions that improved the quality of the paper.
\end{acknowledgments}
%



\end{document}